# TWO-DIMENSIONAL LATTICE GRAVITY AS A SPIN SYSTEM [*]

W. BEIRL, H. MARKUM, J. RIEDLER

*Institut für Kernphysik, Technische Universität Wien, A-1040 Vienna, Austria*

ABSTRACT

Quantum gravity is studied in the path integral formulation applying the Regge calculus. Restricting the quadratic link lengths of the originally triangular lattice the path integral can be transformed to the partition function of a spin system with higher couplings on a Kagomé lattice. Various measures acting as external field are considered. Extensions to matter fields and higher dimensions are discussed.

Thirty years ago Regge developed a discrete description of General Relativity in which space-time is triangulated by a simplicial lattice, the Regge skeleton[1]. Thus, the lattice becomes a dynamical object, with the edge lengths describing the evolution of space-time. Within this scheme the Einstein-Hilbert action for lattice gravity in two dimensions is given by[2]

$$S = \lambda \sum_t A_t - 2k \sum_v \delta_v \;, \tag{1}$$

with $\lambda$ a cosmological constant and $A_t$ the area of triangle $t$. The curvature of the surface is concentrated at the vertices $v$ of the lattice leading to a deficit angle $\delta_v$. According to the Gauss-Bonnet theorem the term proportional to the coupling $k$ in the action is a topological invariant equal to $4\pi$ times the Euler characteristic of the surface. Therefore, it can be dropped for a manifold with the fixed topology of a torus as considered in this paper.

In the path integral formulation a quantization of the above action proceeds by evaluating the expression

$$Z = \int d\mu[q] e^{-S[q]} \;. \tag{2}$$

Unfortunately, a unique prescription for the measure does not exist, however, an appropriate choice proposed in the literature is[3]

$$\int d\mu[q] = \prod_l \int \frac{dq_l}{q_l^m} \;, \tag{3}$$

with $m \in \mathbb{R}$ defining a one-parameter family.

The central idea of this investigation is to transform the path integral to a partition function of a spin system. Therefore, we allow all the squared edge lengths to take two values

$$q_l = 1 + \epsilon \sigma_l \;, \quad 0 \leq \epsilon < 0.6, \quad \sigma_l \in Z_2 \;. \tag{4}$$

[*]Supported in part by FWF under Contract P9522-PHY.

The real parameter $\epsilon$ is restricted to fulfill the Euclidean triangle inequalities for the $q_l$'s. To rewrite the action in terms of $\sigma_l$ we consider a single triangle $t$ consisting of edges with the quadratic lengths $q_1$, $q_2$, $q_l$. Its squared area can be expressed by

$$A_t^2 = \begin{vmatrix} q_1 & \frac{1}{2}(q_1+q_2-q_l) \\ \frac{1}{2}(q_1+q_2-q_l) & q_2 \end{vmatrix}$$
$$= \frac{3}{4} + \frac{1}{2}(\sigma_1+\sigma_2+\sigma_l)\epsilon + \frac{1}{2}(\sigma_1\sigma_2+\sigma_1\sigma_l+\sigma_2\sigma_l - \frac{3}{2})\epsilon^2 \ . \qquad (5)$$

Expanding $\sqrt{A_t^2}$ the series consists only of terms up to $\sigma^3$ since $\sigma_l^2 = 1$. This suggests the following ansatz

$$A_t = c_0(\epsilon) + c_1(\epsilon)(\sigma_1+\sigma_2+\sigma_l) + c_2(\epsilon)(\sigma_1\sigma_2+\sigma_1\sigma_l+\sigma_2\sigma_l) + c_3(\epsilon)\sigma_1\sigma_2\sigma_l \ , \qquad (6)$$

where the coefficients $c_i$ can be obtained by comparing with those in (5). Using (4) the measure (3) can be replaced by

$$\sum_{\sigma_l=\pm 1} \exp[-m\sum_l \ln(1+\epsilon\sigma_l)] = \sum_{\sigma_l=\pm 1} \exp[-N_1 m_0(\epsilon) - \sum_l m_1(\epsilon)\sigma_l] \ , \qquad (7)$$

with $m_0 = -\frac{1}{2}m\epsilon^2 + O(\epsilon^4)$, $m_1 = m(\epsilon + \frac{1}{3}\epsilon^3) + O(\epsilon^5)$ and $N_1$ the total number of links. Inserting (6) and (7) into the partition function (2) yields

$$Z = \sum_{\sigma_l=\pm 1} J \exp\{-\sum_l (2\lambda c_1 + m_1)\sigma_l - \lambda \sum_t [c_2(\sigma_1\sigma_2+\sigma_1\sigma_l+\sigma_2\sigma_l) + c_3\sigma_1\sigma_2\sigma_l]\} \ , \qquad (8)$$

with $J = \exp(-\lambda N_2 c_0 - N_1 m_0)$ and $N_2$ the total number of triangles. Thus, the path integral became the partition function of a system consisting of a spin $\sigma_l$ corresponding to each link $l$, with an external "magnetic field" and with 2- and 3-spin nearest neighbor interactions. Drawing the interactions as lines a Kagomé lattice[4] is obtained (cf. Figure 1). Removing the term linear in $\sigma$ by a convenient choice of measure ($m_1 = -2\lambda c_1$) and neglecting 3-spin couplings we get the partition function of an Ising model on a Kagomé lattice. In the ferromagnetic regime the system shows a $1^{st}$-order phase transition, whereas the antiferromagnetic regime is governed by frustration. Because of positive 2-spin couplings, $c_2 > 0$, the gravitational system is in the antiferromagnetic phase. The 3-spin couplings turn out to be negative, $c_3 < 0$, but not strong enough to remove the frustration.

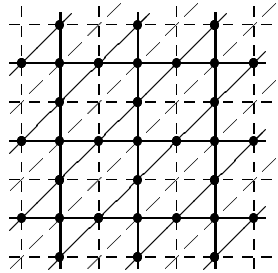

Fig. 1. Original triangular lattice (dashed lines) with the assigned spins (●) and the resulting Kagomé lattice (solid lines).

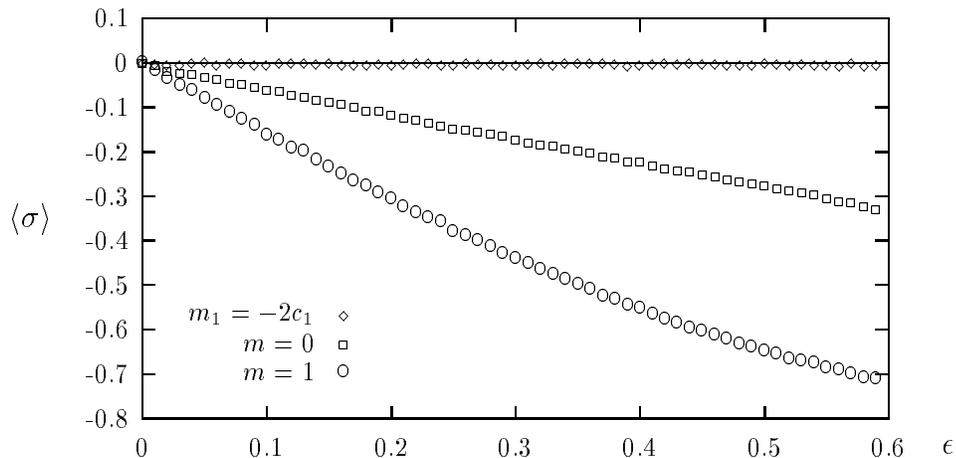

Fig. 2. $\langle\sigma\rangle$ versus $\epsilon$ for $m_1 = -2c_1$ (no external field), $m = 0$ (uniform measure) and $m = 1$ (scale invariant measure). In all cases the cosmological constant is $\lambda = 1$.

We performed numerical simulations with regard to different types of measures. Figure 2 depicts the spin expectation value $\langle\sigma\rangle$ versus $\epsilon$ for the uniform and the scale invariant measure and for the measure leading to a cancellation of the term linear in $\sigma$. It depends on the strength of the measure how strong the frustration is removed. This effect can be seen in both lower curves which favor the occurence of negative spins, corresponding via $q_l = 1 + \epsilon\sigma_l$ to short links reflecting the expected tendency of the lattice to shrink[2].

The motivation for this exploratory study was to approximate quantum gravity by a spin system. It is straightforward to extend this approach from the trivial two-dimensional Regge-Einstein action to scalar fields or physical spin fields. An extension to more than two link lengths can be performed increasing the number of coefficients $c_i$. A generalization of our approach to higher dimensions is possible in principle but becomes more complicated. Notice that in two dimensions only the three links of a triangle are coupled in the action. In three dimensions one is faced with terms of $6^{th}$ order at least, and with additional contributions from the Regge action. The situation is even more complicated in four dimensions where one has to deal with 10 links in each simplex.